\documentclass[a4paper,11pt]{article}

\usepackage{a4wide}
\usepackage{graphics}

\usepackage[%[pdftex%
bookmarks=true%
,bookmarksnumbered=true%
,hypertexnames=false%
,breaklinks=true%
]{hyperref}
\hypersetup{
  pdfauthor = {Djoerd Hiemstra and Vojkan Mihajlovic},
  pdftitle = {A database approach to information retrieval: The remarkable
  relationship between language models and region models},
  pdfsubject = {},
  pdfkeywords = {},
  pdfcreator = {},
  pdfproducer = {}}

\newcommand{\SN}{\mbox{\footnotesize SN}}
\newcommand{\SW}{\mbox{\footnotesize SW}}
\newcommand{\ADJ}{\mbox{\footnotesize ADJ}}

\newcommand{\AND}{\mbox{\footnotesize AND}}
\newcommand{\OR}{\mbox{\footnotesize OR}}
\newcommand{\CONTAINING}{\mbox{\footnotesize CONTAINING}}
\newcommand{\CONTAINEDBY}{\mbox{\footnotesize CONTAINED\_BY}}
\newcommand{\SCALE}{\mbox{\footnotesize SCALE}}

\title{A database approach to information retrieval: The remarkable
  relationship between language models and region models}
\author{Djoerd Hiemstra and Vojkan Mihajlovi{\'c}\\
    University of Twente\\
    Centre for Telematics and Information Technology\\
    P.O. Box 217, 7500 AE Enschede,
    The Netherlands  \\
    \{d.hiemstra,v.mihajlovic\}@utwente.nl
    }
\date{}

\begin{document}

\maketitle

\begin{abstract}
In this report, we unify two quite distinct approaches to information retrieval: region models and language models. Region models were developed for structured document retrieval. They provide a well-defined behaviour as well as a simple query language that allows application developers to rapidly develop applications. Language models are particularly useful to reason about the ranking of search results, and for developing new ranking approaches.
The unified model allows application developers to define complex language modeling approaches as logical queries on a textual database. We show a remarkable one-to-one relationship between region queries and the language models they represent for a wide variety of applications: simple ad-hoc search, cross-language retrieval, video retrieval, and web search.
\end{abstract}

%\vspace{0.1cm}

\section{Introduction}
\label{sec:intro}

The introduction of the relational model by Codd in 1970 \cite{Codd70} marks one of the success stories of computer science. The relational model laid the path for the development of relational database systems: general software tools for management of data with a well-understood and well-defined behaviour. They allow application developers to rapidly develop application programs that are easy to understand, document and teach \cite{Date81}. Indeed, saying ``databases'' is saying ``relational'': Virtually any introductory book or course on databases will teach the basics of the relational data model and SQL\@.

It can be argued that information retrieval is still at the stage where databases were in the 1960's. There is no such thing as an equivalent of the relational model for information retrieval systems. Introductory books and courses on information retrieval \cite{BaezaYates99,Rijsbergen79} will teach the student several information retrieval models --~mostly focusing on different ranking strategies~-- each with its own strengths and weaknesses. Developing a retrieval application or deploying a search engine requires applications to call non-standard application program interfaces (APIs) and use non-standard query languages.

% access path independence & Terrier doc parsing?
% indexing independence & Lemur

As an example, the Terrier system, a research information retrieval system developed by the University of Glasgow \cite{Terrier05}, is based on the so-called divergence of randomness models \cite{Amati02}. Terrier provides APIs for indexing and querying. To use the Terrier indexing API on a non-standard collection (Terrier comes with some fully implemented APIs, for instance for HTML documents), the application developer needs to create an object which implements the collection interface. This will find all the files it has to process, and opens each one to create a document object which identifies which tags (or other byte sequences) act as document delimiters. Applications programs that work with this setup will be logically impaired if the file locations or document format (for instance the XML DTD) need to be changed. Or, in analogy with Codd's \cite{Codd70} analysis of the database systems from the 1960's: The retrieval system does not provide {\em access path independence}.

As another example, the Lemur toolkit \cite{Lemur02} is a research retrieval system that is specifically designed to support research in language modeling \cite{Hiemstra99a,Miller99,Ponte98}. The toolkit supports a broad range of different applications of information retrieval such as ad hoc retrieval, distributed retrieval, cross-language retrieval, etc. Lemur supports at least four different index types, each supporting different kinds of queries. For instance, some indexes include word positions to allow proximity queries, whereas others only allow very basic functionality. Application programs that work with one kind of index might be logically impaired if the index type is changed. In analogy with Codd \cite{Codd70}, the retrieval system does not provide {\em indexing independence}.\footnote{Codd identified one more type of data independence:
    ordering independence. As textual data is inherently ordered we are not
    concerned with ordering independence.}

In the past, we have used systems like Terrier and Lemur to research new applications of information retrieval technology such as cross-language retrieval \cite{Hiemstra99c}, web retrieval \cite{Kraaij02}, and video shot retrieval \cite{Ianeva05}. 
To develop such retrieval approaches, it was necessary to reimplement parts of the existing system: reimplementing APIs, introducing new APIs, introducing new query languages, and even introducing new indexing and storage structures.
In this report, we present a framework that  supports all such approaches by means of a simple yet powerful query language (similar to SQL or relational algebra) that hides the implementation details of retrieval approaches from the application developer. As such, the system provides access path independence and indexing independence.

There have been other attempts to develop approaches to information retrieval that provide data independence. For instance, Schek \cite{Schek80} describes methods for integrating databases and information retrieval systems where application programs and queries are not aware of access paths and indexes. Fuhr \cite{Fuhr99} describes a layered system design for information retrieval systems following the ANSI/SPARC model \cite{Tsichritzis78}, distinguishing a physical (internal) layer, a conceptual layer and an external layer. The system might process queries in several ways, such as directly by an index, or by using an index as a filter with an additional scan of the filtered results. Probabilistic relational algebra or probabilistic Datalog (see \cite{Fuhr96b} for an overview) might serve as conceptual query languages in such systems. An example of a system that implements this approach is HySpirit \cite{Fuhr98}.
In this report we introduce an alternative for probabilistic relational
algebra and probabilistic Datalog that is much closer to existing models of
information retrieval.

\subsection{Region models}
\noindent
~ Motivated by the data independence issues described above, Burkowski \cite{Burkowski92} proposes a mathematical framework which he called the {\em containment model} that operates on sets of contiguous extents. We will call extents {\em regions} in this report, and the model {\em region model}. A region might be a word, a phrase, a text element such as a title, or a complete document. Burkowski's model comes with a small number of basic operators on sets of regions, the most important ones being {\SN} (select narrow) and {\SW} (select wide). A search for chapters containing the word ``databases'' would be expressed as {\tt\small <chapter> {\SW} databases}, and if the application program only needs to put the chapter's title on the screen, the query would be {\tt\small <chapter{\textunderscore}title> {\SN} (<chapter> {\SW} databases)}. In Burkowski's framework, the application program does not know how a text collection and its index facilities are managed. The complexity of the retrieval system is encapsulated in a module that only responds to simple command strings like the ones above. Similar frameworks are introduced by Salminen and Tompa \cite{Salminen92}, Clarke et al. \cite{Clarke95}, Baeza-Yates and Navarro \cite{BaezaYates97}, Consens and Milo \cite{Consens95}, and Jaakkola and Kilpelainen \cite{Jaakkola99}. We will call the models underlying these approaches {\em region models} in this report.

Unlike Codd's relational model for databases, the region models above did not have a big impact on the information retrieval research community, nor on the development of new retrieval systems. The reason for this is quite obvious: region models do not explain in anyway how search results should be ranked. In fact, most region models are not concerned with ranking at all; one might say they --~like the relational model~-- are actually data models instead of information retrieval models. Region model approaches that do address ranking, like Burkowski's model \cite{Burkowski92} and the approach by Masuda et al. \cite{Masuda03}, only include it as an after-thought: Retrieve first, then rank with some standard retrieval model such as a vector space model using {\em tf.idf} weights \cite{Salton88}.

\subsection{Language models}
If anything, an approach to information retrieval has to address the ranking of search results. Ranking is the single most important feature of a search engine, and information retrieval modeling almost exclusively focuses on ranking (see e.g.\ \cite[Chapter 2]{BaezaYates99}).
Traditionally, developing ranking strategies involves engineering, fitting and tuning term weighting approaches to improve experimental results \cite{Salton88}, although there are some notable exceptions, for instance the probabilistic model by Robertson and Sparck-Jones \cite{Robertson76}. A more recent approach that does not require lots of fitting and tuning are statistical language models for information retrieval \cite{Hiemstra99a,Miller99,Ponte98}. Language models assign a probability to a piece of text. They are built for each document: Each document model assigns a probability to a text query, and documents are ranked accordingly. Language models have been applied to a wide variety of retrieval problems, such as simple ad-hoc search \cite{Hiemstra99a,Kamps05,Miller99}, cross-language retrieval \cite{Berger99,Hiemstra99c,Lavrenko02,Xu01}, video retrieval using speech transcripts \cite{Cooke05,Ianeva05}, and web search \cite{Kamps05,Kraaij02,OgilvieCallan04}.
%, without the need for much tuning.
Examples of these applications will be shown in Section~\ref{sec:queries}.

%\subsection{$\!\!\!\!\!\!\!$Unifying region$\,$models and %language$\,$models$\!\!\!\!$}
\subsection{Unifying region models and language models}
In this report we introduce an approach to information retrieval that fully integrates region models and language models. The approach allows application developers to define complex language modeling approaches  as logical region queries on a textual database. We show a remarkable one-to-one relation between region queries and the language models they represent for the four retrieval problems mentioned above: ad-hoc search, cross-language retrieval, video retrieval, and web search. The report is organised as follows. In Section \ref{sec:model} we introduce the combined region/language model. Section \ref{sec:queries}  illustrates the application of the model by relating probability measures to region queries. Finally in Section \ref{sec:conclusions} we present future work and relate the approach to current work on XML query languages and XML database systems.

%\vspace{0.1cm}

\section{A region model for text databases and a query language}
\label{sec:model}
This section briefly introduces the unified region/language model. The definitions closely follow Burkowski's model \cite{Burkowski92}, which we extend with region scores similar to the score region algebra we used for XML information retrieval \cite{List05}.

A {\bf textual database} consists of a finite sequence of {\bf words}
$w_1, w_2, \cdots, w_{n-1}$, where $w_i$ is used to denote the word on position $i$ in the database. Additionally, the textual database consists of a hierarchy of text {\bf elements}. Both words and elements are identified by the word positions in the database. Text elements are sequences of words that have a particular significance in the database. For example, a database with recipes will have text elements ``ingredients'', ``quantities'', ``instructions'', etc., typically marked up as XML\@.

A {\bf scored region} $r$ is defined by two integers $r.start$ and $r.end$ ($1 \le r.start < r.end \le n$), and a float $r.score$
($r.score > 0$).\footnote{We
   intentionally use a notation that is close to that
   of the relational data model; see also Figure \ref{fig:sql}.}
The integers {\em start} and {\em end} represent respectively the position of the first word that belongs to the contiguous region, and the position directly following the last word that belongs to the region. A region might be a text element, but also any other contiguous sequence of words. Note that the region $(i, i+1, s)$ includes one (and only one) word $w_i$ with a score $s$.

Retrieval from the textual database is done with a simple query language consisting of words, elements and five basic {\bf operators}: {\CONTAINING}, {\CONTAINEDBY}, {\SCALE}, {\AND}, and {\OR}\@. The language defines an algebra on sets of scored regions. Unlike Burkowski's model \cite{Burkowski92}, there are no additional constraints on sets of regions. We will now one-by-one define the language primitives in a rather informal way. For convenience, Figure \ref{fig:sql} contains a more formal definition of the operators using SQL\@.

\begin{description}
\item[A word]~ A single word, for example the query {\tt banana}, produces a set of regions $R$, where each region $r \in R$ defines a position of the word in the textual database; $r.start$ being the position on which the word occurs, $r.end = r.start+1$, and $r.score = 1$.
\item[An element]~ A single element, for instance the query {\tt\small <recipe>} produces a set of regions $R$, where each region $r \in R$ is tagged as ``recipe'', $r.start$ being the position of the first word of the XML element, $r.end$ being the position following the last word of the XML element, and $r.score = 1$.
\item[$R_1$ {\CONTAINING} $R_2$]~
The operator {\CONTAINING} takes two sets of regions $R_1$ and $R_2$, and produces the subset of regions from $R_1$ that contain at least one region from $R_2$\@. For instance, the query {\tt\small <recipe> {\CONTAINING} banana} produces all regions tagged as ``recipe'' that contain at least one occurrence of ``banana''. Inspired by language models, each ``recipe'' region is scored by the number of occurrences of ``banana'' in the region, divided by the length of the region (measured as $r.end - r.start$).
Occurrences of ``banana'' are weighted by their length and by their score (of course, in the example query both length = 1 as well as score = 1); see Figure \ref{fig:sql}.
\item[$R_1\,${\CONTAINEDBY}$\,R_2$]
The operator {\CONTAINEDBY} takes two sets of regions $R_1$ and $R_2$, and produces the subset of regions from $R_1$ that are at least contained by one region from $R_2$. For instance, the query {\tt\small <ingredient> {\CONTAINEDBY} <recipe>} produces all ingredients that belong at least to one recipe. If a region from the left-hand side of the expression is nested in more than one region from the right-hand side of the expression, then the scores of those regions are added. This will be used in the next section to express the linear combination of several language models; see Figure \ref{fig:sql}.
\item[$f$ {\SCALE} $R$]~
The operator {\SCALE} takes a float $f$ and a set of regions $R$ and produces all regions from $R$ where each region $r \in R$ is scored as $f \cdot r.score$.
For instance, the query {\tt\small 0.2 {\SCALE} banana} produces the set of regions with the positions of the word ``banana'' all with a region score of
0.2; see Figure \ref{fig:sql}.
\item[$R_1$ {\AND} $R_2$]
The operator {\AND} takes two sets of regions $R_1$ and $R_2$, and produces only those regions that are both in $R_1$ and $R_2$, i.e., the intersection of both sets when ignoring the region scores. Each region in the result is scored
by multiplying its scores in $R_1$ and  $R_2$\@. For instance, the query
{\tt\small (<recipe> {\CONTAINING} banana) \,{\AND}\, (<recipe> \,{\CONTAINING}\, apple)}
produces all regions tagged as ``recipe'' that contain both the word ``banana'' and the word ``apple'', scored by the product of the scores of the respective regions; see Figure \ref{fig:sql}.
\item[$R_1$ {\OR} $R_2$]~
The operator {\OR} takes two sets of regions $R_1$ and $R_2$, and produces those regions that either are in $R_1$, or in $R_2$, i.e., the union of both sets when ignoring the region scores.
%Regions in the result is scored as $R_1.score + R_2.score$.
For instance, the query
{\tt (<recipe> {\CONTAINING} sugar)$\;${\OR}$\;$(<recipe> {\CONTAINING} sweet)}
produces all regions tagged as ``recipe'' that contain either the word ``sugar'' or the word ``sweet'' (or both). Regions keep their score, unless both sets contain the region, in which case the region is scored by adding its scores in $R_1$ and $R_2$; see Figure~\ref{fig:sql}.
\end{description}

\begin{figure}[htb]
\begin{center}
\begin{small}\begin{ttfamily}
\renewcommand{\arraystretch}{0.8}
\begin{tabular}{|l|} \hline
\\
-- R1 CONTAINING R2\\
SELECT R1.start, R1.end, R1.score * SUM((R2.score *\\
~~(R2.end - R2.start)) / (R1.end - R1.start)) AS score\\
FROM R1, R2\\
WHERE R1.start <= R2.start AND R1.end >= R2.end\\
GROUP BY R1.start, R1.end, R1.score\\
\\
-- R1 CONTAINED{\textunderscore}BY R2\\
SELECT$\;$R1.start$\!$,$\;$R1.end$\!$,$\;$R1.score$\;$*$\;$SUM(R2.score)$\;$AS$\;$score\\
FROM R1, R2\\
WHERE R1.start >= R2.start AND R1.end <= R2.end\\
GROUP BY R1.start, R1.end, R1.score\\
\\
-- f SCALE R\\
SELECT R.start, R.end, f * R.score AS score\\
FROM R\\
\\
-- R1 AND R2\\
SELECT R1.start, R1.end, R1.score * R2.score AS score\\
FROM R1, R2\\
WHERE R1.start = R2.start AND R1.end = R2.end\\
\\
-- R1 OR R2\\
SELECT R.start, R.end, SUM(R.score) AS score\\
FROM (SELECT$\;$*$\;$FROM R1 UNION ALL SELECT$\;$*$\;$FROM R2) AS R\\
GROUP BY R.start, R.end\\
\\
\hline
\end{tabular}
\renewcommand{\arraystretch}{1}
\end{ttfamily}
\end{small}
  \caption{Definition of operators in SQL.}\label{fig:sql}
\end{center}
\end{figure}

Figure \ref{fig:sql} contains a definition of the operators using SQL, as a pragmatic means to provide a {\em formal} definition of the region algebra operators without the need to get into specific mathematical notations. So, we show SQL definitions here for convenience, as we assume most readers are familiar with SQL\@. The definitions do not suggest in any way that the system should be implemented on top a relational databases system. We implemented the system --~without the use of SQL~-- on top of MonetDB \cite{List05}, but it might as well be implemented using traditional inverted file indexes on the file system.\footnote{For readers that do want to implement this on top of a relational DBMS, please note
%Note that `f' acts as a host variable in the third SQL statement.
that `R1.end' clashes with the SQL reserved word `END' in practical systems.}

A natural application of the region model, is to support structured queries in an XML information retrieval system. The following query is an example XML information retrieval query formulated in NEXI\@. NEXI \cite{Trotman04} stands for narrowed extended XPath, a query language that restricts XPath \cite{XPath} by only allowing descendent axis steps, and that extends XPath by a special about operator that ranks the selected nodes by their estimated relevance to the query. NEXI is used to evaluate XML retrieval systems in the Initiative for the Evaluation of XML retrieval (INEX) \cite{INEX04}.
Suppose we want to retrieve sections about ``databases'' from articles that mention ``book review'' in either the article title (atl) or the keywords (kwd):
\begin{displaymath}
%\begin{array}{l}
 \mbox{\tt\small //article[about(.//(atl|kwd), book review)]//sec[about(., databases)]}\\
% \mbox{\tt\small }
%\end{array}
\end{displaymath}
This can be formulated as follows as a region query:
\begin{displaymath}
\begin{array}{l}
 \!\!\!\mbox{\tt\small
   (<sec> $\!${\CONTAINING}$\!$ databases) $\!${\CONTAINEDBY}$\!$
      (<article> $\!${\CONTAINING}$\!$}\\
 \mbox{\tt\small ~~~~ (((<atl> $\!${\OR}$\!$ <kwd>)$\!$ {\CONTAINING}$\!$ book) $\!${\CONTAINING}$\!$ review))
  }
\end{array}
\end{displaymath}
This approach is followed with success in INEX by the TIJAH system \cite{List05,Mihajlovic05}. The expression defines a ranking of the selected nodes. Rewriting the NEXI query to the region expression is not trivial, but relatively easy: TIJAH has a NEXI to region query parser.

In the next section we show the relationship between language modeling ranking definitions and region queries, similar to the relationship between NEXI queries and the region queries.

\vspace{0.24cm}

\section{Logical queries for complex retrieval tasks}
\label{sec:queries}

\subsection{\label{sec:lmsimple}The simplest unigram language model}
As said in the introduction, language models form a general approach to define ranking formulas for retrieval applications. A language model is assigned to every document. The language model of the document defines the probability that the document `generates' the query. Documents are ranked by this probability. The simplest language modeling approach to information retrieval would be defined by Equation \ref{eq:lmsimple}.
\begin{equation}
  P(T_1, T_2, \cdots, T_l | D) = \prod_{i=1}^{l} P(T_i | D)
  \label{eq:lmsimple}
\end{equation}

\noindent
It defines the probability of a query of length $l$ given a document $D$ as the product of the probabilities of each term $T_i$ $(1 \le i \le l)$ given $D$\@.
A language model that takes a simple product of terms, i.e., a model that assumes that the probability of one term given a document does not depend on other terms, is called a {\em unigram} language model.
To make this work, we have to define the basic probability measure $P(T|D)$; typically, it would be defined as the number of occurrences of the term $T$ in the document $D$, divided by the total number of terms in the document $D$.
For a practical query, say, {\em retrieve all documents about ``db'' and ``ir''}, we would instantiate Equation \ref{eq:lmsimple} as follows:
\begin{equation}
 P(T_1\!=\!\mbox{\tt\small db}, T_2 = \mbox{\tt\small ir} | D)
                   \; = \; P(T_1\!=\!\mbox{\tt\small db} | D)
                     \; \cdot \;
                      P(T_2\!=\!\mbox{\tt\small ir} | D)
\label{eq:insimple}
\end{equation}

\noindent
The right-hand side of the equation corresponds to the following region expression.
\begin{equation}
 \hspace{1.24cm}\mbox{\tt\small
  (<doc> $\!${\CONTAINING}$\!$ db)
  $\!${\AND}$\!$
   (<doc> $\!${\CONTAINING}$\!$ ir)
  }
  \label{eq:regsimple}
\end{equation}

\noindent
This can be shown as follows: The region expression
  {\tt\small (<doc> {\CONTAINING} db)}
produces all documents ranked according to $P(T = \mbox{\tt\small db} | D)$, i.e., all regions tagged as {\tt\small <doc>}, ranked by the number of occurrences of {\tt\small db} in those regions. Similarly,
  {\tt\small (<doc> {\CONTAINING} ir)}
produces all documents ranked according to $P(T = \mbox{\tt\small ir} | D)$.
Finally, the operator {\tt\small \AND} results in the regions tagged as {\tt\small <doc>} that are in both operand sets. The score of the result regions is defined as the product of the scores of the same regions in the operands.
Here, and in the remaining examples in this section, we assume that {\tt\small <doc>} regions do not nest inside each other.

We claim that there is a trivial way to rewrite the right-hand side of Equation \ref{eq:insimple} to Equation \ref{eq:regsimple} while preserving the outcome. This can be shown by simply replacing
  $P(x|y)$ by {\tt\small (y {\CONTAINING} x)},
and the multiplication in Equation \ref{eq:insimple} by {\tt\small \AND}\@.
Regions that are assigned zero probability by the probability measure of Equation~\ref{eq:insimple} are not retrieved by the region expression of Equation~\ref{eq:regsimple}.
So, the region expression selects all $y$ for which $P(x|y) > 0$. If the probability measure assigns zero probability to a region then this implies that the corresponding region expression will not retrieve it; and, if a region is not retrieved by a region expression then this implies that its corresponding probability function assigns zero probability to it.

\subsection{Linear interpolation smoothing}
The simple language model presented in the previous section assigns zero probability to a document unless it contains all query terms. So, if none of the documents contains all terms, the system does not retrieve anything. This behaviour will be appropriate for many practical applications. In fact, it is the default behaviour of web search engines like Google and Yahoo.

For other applications, it might be undesirable to have empty results. When searching collections that are significantly smaller than the web, it is likely that precise queries will not retrieve anything. In practice, language modeling approaches therefore use a technique called ``smoothing'', i.e., some probability mass is assigned to terms that do not occur in the document. The standard language modeling approach uses a mixture of the document model $P(T_i|D)$ with a general collection model $P(T_i|C)$
\cite{Berger99,Hiemstra99a,Lafferty01,Miller99,Ng00,Song99}, called {\em linear interpolation smoothing}.
\begin{equation}
  P(T_1, T_2, \cdots, T_l | D) =
    \prod_{i=1}^{l} ((1\!-\!\lambda) P(T_i|C) +  \lambda P(T_i | D))
    \label{eq:lmsmooth}
\end{equation}

\noindent
The document model $P(T_i|D)$ assigns zero probability to
terms that do not occur in the document $D$, but the collection model $P(T_i|C)$ assigns some probability to any term that occurs somewhere in the collection. The collection model probabilities are defined similar to the document model probabilities as: The number of occurrences of the term $T$ in the total collection $C$, divided by the total number of terms in the collection $C$.
The approach needs a parameter $\lambda$ $(0 < \lambda < 1)$ which is set empirically.

For our example query, we need some value for $\lambda$ to instantiate Equation \ref{eq:lmsmooth}. Suppose we decide $\lambda=0.8$, then we would rank documents according to:
\begin{equation}
\begin{array}{l}
P(T=\mbox{\tt\small db} , T=\mbox{\tt\small ir} | D) \; = \\
\;(0.2 \!\cdot\! P(T_1\!=\!\mbox{\tt\small db} | C) + 0.8 \!\cdot\! P(T_1\!=\!\mbox{\tt\small db} | D))\\
                     \; \cdot \;\\
\; (0.2 \!\cdot\! P(T_2\!=\!\mbox{\tt\small ir} | C) + 0.8 \!\cdot\! P(T_2\!=\!\mbox{\tt\small ir} | D))
\hspace{1.8cm}
\label{eq:insmooth}
\end{array}
\end{equation}

\noindent
The equation corresponds to the following region expression, where the text element {\tt\small <root>} corresponds to the collection root, i.e., the whole database.
\begin{equation}
\begin{array}{l}
 \mbox{\tt\small
 (<doc> {\CONTAINEDBY}}\\
 \mbox{\tt\small ~((0.2 {\SCALE} (<root> $\!${\CONTAINING}$\!$ db))
    $\!${\OR}$\!$ (0.8 {\SCALE} (<doc> $\!${\CONTAINING}$\!$ db))))}\\
  \mbox{\tt\small {\AND}}\\
 \mbox{\tt\small (<doc> {\CONTAINEDBY}}\\
 \mbox{\tt\small ~((0.2 {\SCALE} (<root> $\!${\CONTAINING}$\!$ ir))
   $\!${\OR}$\!$ (0.8 {\SCALE} (<doc> $\!${\CONTAINING}$\!$ ir))))
  }\!\!
\end{array}\label{eq:regsmooth}
\end{equation}

\noindent
This can be shown as follows: The region expression
  {\tt\small (<root> $\!${\CONTAINING}$\!$ db)}
results in a set with the single region {\tt\small <root>} with a score equal to the number of occurrences of {\tt\small db} in {\tt\small <root>}, i.e., $P(T|C)$. The {\SCALE} operator will multiply the region with 0.2; and the {\OR} will union the region with all document regions (with scores $P(T|D)$ as in the previous section), multiplied with 0.8 by the {\SCALE} operator. Note, that the {\OR} operator will not actually add $0.2 \cdot P(T\!=\!\mbox{\tt\small db} | C)$ to $0.8 \cdot P(T\!=\!\mbox{\tt\small db} | D)$: This will be done by the {\CONTAINEDBY} operator: every document region on the left-hand side of this operator matches (because every document region is contained by the collection root). Document regions that are in the set  {\tt\small 0.8 {\SCALE} (<doc> $\!${\CONTAINING}$\!$ db)} will get as their final score: $0.2 \cdot P(T\!=\!\mbox{\tt\small db} | C) + 0.8 \cdot P(T\!=\!\mbox{\tt\small db} | D)$; the others will get: $0.2 \!\cdot\! P(T\!=\!\mbox{\tt\small db} | C)$\@. The same line of reasoning can be done for the part with the term {\tt\small ir}. Finally, the {\AND} operator combines both parts of the query as in the previous section.

Again, we claim there is a trivial way to rewrite the right-hand side of Equation \ref{eq:insmooth} to Equation \ref{eq:regsmooth}. This can be shown by simply replacing
  $P(x|y)$ by {\tt\small (y {\CONTAINING} x)},
the multiplication operator `$\cdot$' by {\tt\small \AND} if both operands are regions, or by {\tt\small \SCALE} if the first operand is a number;
the addition operator `$+$' by {\tt\small \OR}, and by putting
  ``z {\CONTAINEDBY}'' in front of the expression, where $z$ defines the elements that need to be retrieved.

It might be argued that this very last step --~``putting {\CONTAINEDBY} in front''~-- is not a trivial step, and we did not use it in the previous section. However, we might as well use it in the previous section: It is easy to show that
  {\tt\small (<doc> {\CONTAINING} db) {\AND} (<doc> {\CONTAINING} ir)}
produces the same regions, with the exact same scores as
{\tt\small (<doc> $\!${\CONTAINEDBY}$\!$ (<doc> $\!${\CONTAINING}$\!$ db)) $\!${\AND}$\!$ (<doc> $\!${\CONTAINEDBY}$\!$ (<doc> $\!${\CONTAINING}$\!$ ir))},
because the elements on the left-hand side of both {\CONTAINEDBY} operators all have unit score, and because elements on the left-hand side are nested in at most one region from the right-hand side of the {\CONTAINEDBY} operator. So, the general procedure that rewrites probability measures to region expressions should use the {\CONTAINEDBY} operator for every query term.
Equivalences between region expressions will be addressed briefly in Section \ref{sec:optimization}.

\subsection{Video shot retrieval using speech transcripts}
Now that we showed linear interpolation smoothing, it is easy to generalise this to any linear combination of language models. Such models have been quite successful in spoken document retrieval for retrieving video shots \cite{Cooke05,Ianeva05}, where videos are modeled as sequences of scenes, each consisting of sequences of shots. The language model mixes four different levels of the video hierarchy: shots, scenes, complete videos and the total collection as:
\begin{equation}
\vspace{0.1cm}
\begin{array}{l}
  P(T_1, T_2,\! \cdots\!, T_l | Shot) \,=\,\\
    \;\;\;{\displaystyle
    \prod_{i=1}^{l} (\alpha P(T_i|C) + \beta P(T_i | Video) +
       \gamma P(T_i | Scene) + \delta P(T_i | Shot)) }
\end{array}
\label{eq:lmvideo}
\end{equation}

\noindent
where $\alpha + \beta + \gamma + \delta = 1$. The main idea behind this approach is that a good shot contains the query terms, and is part of a scene that contains the query terms, which is part of a video that contains even more of the query  terms.
%Similar approaches have been successful for passage retrieval \cite{allan}.
%
Suppose we are looking for the exact shots in a collection of videos where a  knight says ``ni'',\footnote{From the movie ``Monty Python and the Holy Grail''} and we take $\alpha=0.18$, $\beta=0.02$, $\gamma=0.4$, and $\delta=0.4$ then the shots would be ranked according to:
\begin{equation}
\vspace{0.1cm}
\begin{array}{l}
  P(T\!=\!\mbox{\tt\small ni} | Shot) \;\;= \\
  \hspace{0.9cm} (0.18 \!\cdot\! P(T\!=\!\mbox{\tt\small ni}|C)  \;+\;
               0.02 \!\cdot\! P(T\!=\!\mbox{\tt\small ni} | \mathit{Video}) \\
  \hspace{0.9cm}\;\;\; +\, 0.4 \!\cdot\! P(T\!=\!\mbox{\tt\small ni} | \mathit{Scene})  \,+\,
               0.4 \!\cdot\! P(T\!=\!\mbox{\tt\small ni} | \mathit{Shot}))
\end{array}
\label{eq:invideo}
\end{equation}

\noindent
which corresponds to the following region expression.
\begin{equation}
\vspace{0.1cm}
\begin{array}{l}
 \!\!\!\!\mbox{\tt\small
 $\!\!\!\!$<shot> {\CONTAINEDBY}}\\
 \mbox{\tt\small ~
   ((0.18 $\!${\SCALE}$\!$ (<root> $\!${\CONTAINING}$\!$ ni)) $\!${\OR}$\!$
     (0.02 $\!${\SCALE}$\!$ (<video> $\!${\CONTAINING}$\!$ ni))}\!\!\!\!\\
 \mbox{\tt\small ~~~
    $\!${\OR}$\!$
     (0.4 $\!${\SCALE}$\!$ (<scene> $\!${\CONTAINING}$\!$ ni)) $\!${\OR}$\!$
     (0.4 $\!${\SCALE}$\!$ (<shot> $\!${\CONTAINING}$\!$ ni)))
  }\!\!\!\!
\end{array}
\label{eq:regvideo}
\end{equation}

\noindent
Showing that the region expression of Equation \ref{eq:regvideo} retrieves and ranks video shots according to Equation \ref{eq:invideo} is done as in the previous section.

\vspace{0.1cm}

\subsection{Web retrieval with page priors}
For web retrieval, non-content information like the number of hyperlinks pointing to a web page, or the form of the URL are good indicators of the importance of a page. Such approaches can be modeled by so-called document priors $P(D)$ that do not depend on the query \cite{Kamps05,Kraaij02,OgilvieCallan04}. Document priors are calculated once for the  entire collection, stored in the system and then used to enhance retrieval results for every query. A good example of such an approach is Google's PageRank algorithm \cite{BrinPage98}.

Document priors are motivated as follows.
Instead of ranking documents by the probability that they generate the query, it makes more sense to rank them by $P(D | T_1, T_2,\!\cdots\!, T_l)$: The probability that $D$ is relevant given the query $T_1, T_2,\!\cdots\!, T_l$ of length $l$. According to Bayes' rule:
\begin{equation}
\vspace{0.1cm}
\begin{array}{rcl}
 P(D | T_1, T_2,\!\cdots\!, T_l) & \!=\! &
    \!{\displaystyle \frac{P(D) \cdot P(T_1, T_2,\!\cdots\!, T_l | D)}
                        {P(T_1, T_2,\!\cdots\!, T_l)}}\\
 & \!\propto\! & \!{\displaystyle P(D)\cdot \prod_{i=1}^{l} P(T_i | D)}\\
\end{array}
\vspace{0.1cm}
\label{eq:lmprior}
\end{equation}
The denominator, $P(T_1, T_2, \cdots, T_l)$, does not depend on $D$ and can therefore be dropped, but document prior, $P(D)$, cannot be dropped unless it is uniformly distributed over all documents.
Suppose we are looking for the entry page of Google. Documents will be ranked as follows.
\begin{equation}
\vspace{0.1cm}
\begin{array}{rcl}
 P(D | T\!=\! \mbox{\tt\small google})  & \!\propto\! & \!{\displaystyle
     P(D)\cdot P(T\!=\! \mbox{\tt\small google}| D)}\\
\end{array}
\vspace{0.1cm}
\label{eq:inprior}
\end{equation}
To follow this approach, the system needs to have some means to store text elements with their prior probability. Suppose an application program calculated the PageRank of each crawled web page resulting in probabilities $P(D)$ (or any number proportional to the probabilities, see \cite{BrinPage98}) for each document region, which is stored as {\tt\small \$PageRank}. The dollar sign is used to denote a region set that is stored by the system for later use. The set is used in the query as follows.
\begin{equation}
 \vspace{0.1cm}
 \hspace{1.6cm}\mbox{\tt\small
   \$PageRank
  $\!${\AND}$\!$
   (<doc> $\!${\CONTAINING}$\!$ google)
  }
\label{eq:regprior}
\vspace{0.1cm}
\end{equation}
We believe the correspondence between Equation \ref{eq:inprior} and \ref{eq:regprior} is obvious. As before, the query
   {\tt\small \$PageRank
  $\!${\AND}$\!$
   (<doc> {\CONTAINEDBY} (<doc> $\!${\CONTAINING}$\!$ google))}
would be a more general query that produces the exact same results.

\vspace{0.2cm}

\subsection{Cross-language information retrieval}
In cross-language information retrieval, a collection in one language, e.g.\ English, is searched by querying it in another language, e.g.\ Dutch. A language modeling approach to cross-language retrieval ranks documents by the probability  $P(S_1,S_2,\cdots,S_l|D)$ of generating a Dutch query $S_1,S_2,\cdots,S_l$ of length $l$ from the English document $D$\@. This is modeled by the following procedure: first an English word $T$ is generated from a document with probability $P(T|D)$, then the English term is translated to Dutch independently from the document it was generated from, so with probability $P(S|T)$, resulting in \cite{Berger99,Hiemstra99c,Xu01}:

\begin{equation}
  P(S_1,S_2,\cdots,S_l|D) \;=\; \prod_{i=1}^l \sum_{j=1}^V (P(S_i|T_j)P(T_j|D))
\label{eq:lmclir}
\end{equation}
where $P(T_j|D)$ is again the document language model, and $P(S_i|T_j)$ is a translation model defining the probabilities of the source language words (for instance Dutch in case of a Dutch query) given the target language words (English if the collection being searched is English), and where $V$ is the size of the target language vocabulary. Such a model is used as follows: Given a Dutch query $S_1,S_2,\cdots,S_l$, every word might have several possible translations in English. Suppose we want to use the Dutch query
{\tt\small gebroken hart} (English: ``broken heart'') to search for English documents. The application program would consult its dictionary to determine that there are two possible English translations for the Dutch word ``gebroken'': ``broken'' and ``fractured''. The probability of translating ``broken'' to ``gebroken'', i.e.\
  $P(S=\mbox{\tt\small gebroken} | T =\mbox{\tt\small broken})$
might be estimated as 1.0, for instance because from example texts we know that the English word ``broken'' is always translated to ``gebroken''; and the probability of translating ``fractured'' to ``gebroken'', i.e.\
  $P(S=\mbox{\tt\small gebroken} | T =\mbox{\tt\small fractured})$
might be estimated as 0.2 (note that the two probabilities do not need to sum up to 1). In this case, an instantiation of Equation \ref{eq:lmclir} would be:

\begin{equation}
\!\!\!\!\begin{array}{l}
 P(S_1\!=\!\mbox{\tt\small gebroken}, S_2\!=\!\mbox{\tt\small hart}|D) \;=\\
  \;\; (1.0 \cdot P(T_1\!=\!\mbox{\tt\small broken}|D) \,+\,
     0.2 \cdot P(T_1\!=\!\mbox{\tt\small fractured}|D))\\
  \;\; \cdot \\
  \;\;(0.5 \cdot P(T_2\!=\!\mbox{\tt\small heart}|D) \,+\,
    0.1 \cdot P(T_2\!=\!\mbox{\tt\small ticker}|D)) \\
\end{array}\!\!\!\!
\label{eq:inclir}
\end{equation}

\noindent
So, the sum over the whole target language vocabulary will in practice be a sum over the possible translations only (those for which $P(S|T)>0$).
The probability function corresponds to the following region expression.
\begin{equation}
\vspace{0.1cm}
\!\!\!\!\begin{array}{l}
 \mbox{\tt\small
    ((1.0 $\!${\SCALE}$\!$ (<doc> $\!${\CONTAINING}$\!$ broken)) $\!${\OR}$\!$
     (0.2 $\!${\SCALE}$\!$ (<doc> $\!${\CONTAINING}$\!$ fractured)))}\\
 \mbox{\tt\small {\AND}}\\
 \mbox{\tt\small ((0.5 $\!${\SCALE}$\!$ (<doc> $\!${\CONTAINING}$\!$ heart))
   $\!${\OR}$\!$
    (0.1 $\!${\SCALE}$\!$ (<doc> $\!${\CONTAINING}$\!$ ticker)))
  }
\end{array}\!\!\!\!
\vspace{0.1cm}
\label{eq:regclir}
\end{equation}

\noindent
Equation \ref{eq:regclir} can be generated from \ref{eq:inclir} as shown in the previous sections.

\vspace{0.36cm}

\section{Discussion, open issues and future work}
\label{sec:conclusions}
In this report, we presented a unified region model / language model approach and showed its expressiveness for a wide range of applications of language modeling: ad-hoc retrieval, smoothing, video retrieval, web search and cross-language retrieval. In the past, we have developed separate prototype retrieval systems for these approaches. Developing these prototype systems meant we had to reimplement parts of our system: reimplementing APIs, introducing new APIs, introducing new query languages, introducing new indexes, introducing new storage structures, etc. This report shows that such approaches can be supported by a single retrieval system that responds to a simple query language that hides implementation details of information retrieval approaches from the application developer.

The relationship between the region queries and the language modeling probability functions might seem trivial because we ``hard-wired'' the language modeling probability definition in the {\CONTAINING} operator,
 but we believe it is  remarkable: Note that the language modeling probability functions are arithmetic expressions that define the probability of a single document $D$\@. However, the region queries are algebraic expressions for processing {\em sets} of documents (regions) instead of single documents. Since the region query language forms a ``bulk algebra'', experiences from relational database system design can be used to develop efficient implementations of such a system, possibly up to a point where applications run as fast as, or possibly even faster than, the dedicated prototypes we developed in the past.

\vspace{0.1cm}
\subsection{\label{sec:optimization}Query optimization}
The queries presented in Section \ref{sec:model} are close to the language modeling probability functions. However, there exist alternative expressions of the queries that produce equivalent results but that might be easier to process by the system. Based on a study into equivalence relations for region models \cite{Mihajlovic04}, we conjecture that the following expressions are alternatives for the expressions presented in Section \ref{sec:model}:
{\tt\small (<doc> $\!${\CONTAINING}$\!$ db) $\!${\CONTAINING}$\!$ ir}
is an alternative for Equation  \ref{eq:regsimple};
{\tt\small ~(<doc> {\CONTAINEDBY} (((0.2 {\SCALE} <root>) {\OR} (0.8 {\SCALE} <doc>)) {\CONTAINING} db)) {\CONTAINEDBY} (((0.2 {\SCALE} <root>) {\OR} (0.8 {\SCALE} <doc>)) {\CONTAINING} ir)}
is an alternative for Equation \ref{eq:regsmooth};~
{\tt\small <shot> {\CONTAINEDBY}
   (((0.18 $\!${\SCALE}$\!$ <root>) {\OR}
     (0.02 $\!${\SCALE}$\!$ <video>) {\OR}
     (0.4 $\!${\SCALE}$\!$ <scene>) {\OR}
     (0.4 $\!${\SCALE} <shot>)) {\CONTAINING} ni)}
~is an alternative for Equation \ref{eq:regvideo};
{\tt\small \$PageRank $\!${\CONTAINING}$\!$ google}
is an alternative for Equation \ref{eq:regprior}; finally
{\tt\small (<doc> {\CONTAINING}
   (broken {\OR} (0.2 {\SCALE} fractured))) $\!${\CONTAINING}$\!$
   ((0.5 $\!${\SCALE}$\!$ heart) {\OR} (0.1 $\!${\SCALE}$\!$ ticker))}
is an alternative for Equation \ref{eq:regclir}.

Additionally, query optimization would involve choosing concrete evaluation methods attached to each operation, estimating the costs of each method, and choosing the fastest plan. Ram{\'i}rez and De Vries \cite{Ramirez04} present preliminary results.

\vspace{0.1cm}
\subsection{Towards existing XML query languages}
It can be argued that region models are simple predecessors of models underlying XML query languages like XPath \cite{XPath} and XQuery \cite{XQuery}. That is, operators like {\CONTAINEDBY} and {\CONTAINING} can be seen as ancestor and descendent axis steps, as well as the function {\tt\small fn:contains} in XPath. It would be relatively easy to add other XPath axis steps to the query language if we specify how regions are nested, for instance by requiring that a region has a
{\em level} (the depth in the XML tree) as well as a start, end, and score.

XML and its subsequent standards like XPath and XQuery have initiated a lot of research into XML database systems with dedicated workshops and symposia like
DataX \cite{DataX04} and XSym \cite{XSym04}. Our implementation of the region approach is quite similar to implementations of XML databases that use relational database technology and a numbering of the XML nodes \cite{Tatrinov02}. Interestingly, the word positions that belong to the region start and region end of an XML element are respectively in pre-order and post-order as in the XML database implementation proposed by Grust \cite{Grust02}. Our prototype system TIJAH uses part of the code of the PathFinder XML database system \cite{PathFinder05}. In the future, both systems might be integrated following the XQuery full-text standard \cite{XQueryFullText05,Yahia04}.

\vspace{0.1cm}
\subsection{Towards new applications of XML}
Some people have argued that existing XML query languages like XPath \cite{XPath} and XQuery \cite{XQuery} are too powerful for simple XML information retrieval functionality \cite{Trotman04}. Others have argued that existing query languages are not powerful enough. For instance Ogilvie \cite{Ogilvie04} illustrates a system that answers queries like ``Who killed Abraham Lincoln'' by a query that returns those {\tt\small <person>} elements that directly precede the word {\tt\small killed}, which directly precedes another {\tt\small <person>} element containing {\tt\small lincoln}. Such a query would be hard, if not impossible, to express in existing XML query languages. A solution might be the introduction of a special gluing operator in our region model approach, let's call it {\ADJ} for ``adjacent'', which can glue regions to form bigger regions. Such an operator might be used for phrases, but also to glue for instance two paragraphs together to form a region that spans two paragraphs. We have implemented such a gluing operator in our video retrieval system that, lacking a reliable scene detector, glues adjacent shots together to represent a scene \cite{Ianeva05}.

\vspace{0.1cm}
\subsection{Beyond XML}
Ogilvie \cite{Ogilvie04} also makes a case for allowing several hierarchies of possibly overlapping elements which combined would no longer form a tree. This need is illustrated as well by Burkowski \cite{Burkowski92}, by people studying the bible \cite{DeRose04}, and it is picked up by several initiatives to extend XML \cite{LMNL04,GODDAG99}. The region approach described here would support querying of such representations quite naturally.

\section*{Acknowledgements}
Djoerd Hiemstra was supported by the Dutch BSIK program MultimediaN: Semantic Multimedia Access.
Voj\-kan Mihajlovi{\'c} was supported by the Netherlands
Organisation for Scientific Research (NWO project 612.061.210).
We like to thank Henk Ernst Blok for fruitful discussions on region algebras, and Maarten Fokkinga and Thijs Westerveld (CWI, Amsterdam) for helpful comments on the report.

%\bibliography{all,lmregions}

\end{document}